\documentclass{moriond}

\newcommand{\vev}[1]{\langle {#1} \rangle}
\newcommand{\lsim}{\lesssim}
\newcommand{\gsim}{\gtrsim}
\newcommand{\eq}[1]{Eq.~(\ref{#1})}
\newcommand{\ord}[1]{\mathcal{O}{(#1)}}

\usepackage{amssymb}

\def\be{\begin{equation}}
\def\ee{\end{equation}}
\def\bea{\begin{eqnarray}}
\def\eea{\end{eqnarray}}

\newcommand{\Photo}{\includegraphics[height=35mm]{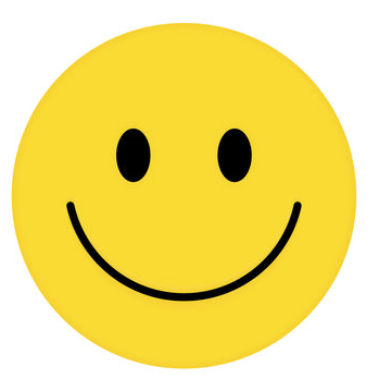}}

\begin{document}
\vspace*{4cm}
\title{DESTABILIZING MATTER THROUGH A LONG-RANGE FORCE
}

\author{HOOMAN DAVOUDIASL}

\address{High Energy Theory Group, Physics Department \\ Brookhaven National Laboratory,
	Upton, NY 11973, USA}

\maketitle\abstracts{We consider a long-range force, mediated by an ultralight scalar, which can give rise to violation of baryon 
	number.  This would lead to very different lifetimes for nucleons in different astrophysical environments.  Possible signals of 
	this scenario include a flux of $\ord{\rm 10~MeV}$ solar neutrinos or anomalous heating of old neutron stars; we find the latter to yield 
	the strongest current bounds, which could be improved in the coming years.  The ultralight scalar employed here can potentially be a good 
	dark matter candidate.}

\section{Introduction}

The following is based on a presentation by the author at the EW Rencontres de Moriond 2024, La Thuile, Italy, March 24-31, 2024.  That presentation provided a summary of the results reported in Ref.~\cite{Davoudiasl:2023peu}, where a full list of relevant references can be found.

The extremely stringent bounds on proton decay can be understood as a consequence of an accidental symmetry that preserves baryon number $B$ in the Standard Model.  Any $B$ violating effect is then due to higher dimension operators suppressed by large masses well above the weak scale.  For example, $B$ violation via proton decay $p\to \pi^0 \ell^+$, where $\pi^0$ denotes the neutral pion, can be mediated by the dimension-6 operator   
\be
O_6=\frac{(u u d\, \ell)_R}{M^2}\,,
\label{O6}
\ee     
where $u$ and $d$ are the up and down quarks, respectively, and $\ell=e,\mu$ is the electron or the muon.  In \eq{O6}, $R$ denotes right-handed fermions.  The partial lifetime bound $\tau > 1.6\, (0.77)\times 10^{34}$~yr, for $e^+$ $(\mu^+)$, at 90\% confidence level \cite{ParticleDataGroup:2022pth} implies that $M\gsim 10^{16}$~GeV, close to the scale of grand unified theories.  

The presence of dark matter (DM) is compelling evidence for new physics beyond the SM.  In principle, DM may belong to a new sector that includes light states.  If $B$ violating operators include such light states, they can change the standard proton decay expectations.  Here, we consider an extreme limit where the $B$ violating operator above is augmented by the addition of an ultralight scalar $\phi$, resulting in the dimension-7 operator
\be
O_7 = \frac{\phi\,(u u d\, \ell)_R}{\Lambda^3}\,,
\label{O7}
\ee
where $\Lambda$ is a new scale set by the ultraviolet theory. We will focus on this operator, as an example to demonstrate the main idea.  

\section{Setup and Formalism}

We will set $m_\phi = 10^{-16}$~eV, corresponding to a Compton wavelength of order the Solar radius $R_\odot \approx 7.0 \times 10^5$~km.  The ultralight scalar is assumed to interact with nucleons with coupling $g_N$ according to   
\be
g_N \phi \bar N\, N\,,
\label{phiNN}
\ee
where $N=p, n$ is a nucleon; $n$ denotes a neutron.  Recent constraints yield $g_N \lsim 8.0\times 10^{-25}$, at $2\sigma$  \cite{Fayet:2017pdp,MICROSCOPE:2022doy}.  Due to the coupling above and the long Compton wavelength of $\phi$, astronomical objects can source a potentially significant background $\vev{\phi}$ which affects the strength of $B$ violation via the operator $O_7$ above.\footnote{This makes $\vev{\phi}$ effectively a spacetime dependent Wilson coefficient for the $B$ violating operator $O_6$ in \eq{O6}.}

We use chiral perturbation theory \cite{Claudson:1981gh} to calculate various decay rates for nucleons, resulting from the operator in \eq{O7}.  For $B$ preserving interactions we find
\be
{\cal L}_{\rm P} = \left[\frac{(3F-D)}{2\sqrt{3}f_\pi}\,\partial_\mu \eta +
\frac{(D+F)}{2 f_\pi}\,\partial_\mu \pi^0\right] \bar p \gamma^\mu \gamma_5 p 
+\frac{(D+F)}{\sqrt{2}f_\pi}\,\partial_\mu \pi^+ \, \bar p \gamma^\mu \gamma_5 n + \ldots\,, 
\label{LP}
\ee      
while $B$ violating interactions are given by 
\be
{\cal L}_{\rm V} = \frac{\beta}{\Lambda^3}\, \phi\, \left[\overline{e^c_R}\, p_R -  
\frac{i}{2 f_\pi} (\sqrt{3} \eta + \pi^0)\overline{e^c_R} \,p_R \right]  
- \frac{\beta}{\Lambda^3}\, \phi\,
\left[\frac{i}{\sqrt{2} f_\pi} \pi^+ \overline{e^c_R} \,n_R\right]
+ {\rm \small H.C.}
\label{LV}
\ee
In the above, $D=0.80$, $F=0.47$, and $\beta=0.012 \pm 0.0026$~GeV$^3$ \cite{Aoki:2008ku}; $f_\pi \approx 92$~MeV is the pion decay constant.
   
We shall focus on 2-body decays (since they are typically dominant) and ignore the positron mass.  
For $p \to {\cal M} e^+$, where ${\cal M} = \pi^0, \eta$ is a meson with mass $m_{\cal M}$, we have 
\be
\Gamma(p \to {\cal M} e^+) = \frac{\lambda_{\cal M}^2}{32 \pi}\,m_p \left(1- \frac{m_{\cal M}^2}{m_p^2}\right)^2\,,
\label{GamM}
\ee
with 
\be
\lambda_\pi \equiv \frac{(D+F+1)\mu}{2 f_\pi} \quad ; \quad \lambda_\eta \equiv \frac{(3F - D + 3)\mu}{2 \sqrt{3} \,f_\pi}\,,
\label{lam-meson}
\ee
$\mu = \kappa \vev{\phi}$, and $\kappa \equiv \beta/\Lambda^3$.    
For the neutron, we get  
\be
\Gamma(n\to \pi^- e^+) = \frac{\lambda_\pi^2}{16 \pi}\,m_n \left(1- \frac{m_{\pi^-}^2}{m_n^2}\right)^2.
\label{neutron-decay}
\ee
In the above expressions, $m_p$ and $m_n$ are the proton and neutron masses, respectively.   

\section{Constraints}

We will start with ``local" constraints, {\it i.e.} from terrestrial experiments and measurements of Solar neutrinos.  The nucleons in the Earth can source a background value 
\be
\vev{\phi_\oplus} \approx -\frac{g_N (M_\oplus/m_N)}{4 \pi \,R_\oplus}\,,
\label{phi-Earth}
\ee 
where the Earth mass $M_\oplus \approx 6.0\times 10^{27}$~g, $m_N\approx 0.94$~GeV is a nucleon mass, and the Earth radius is $R_\oplus \approx 6400~{\rm km} \approx (3\times 10^{-14}~{\rm eV})^{-1}$.  We can employ this background value for our reference parameters and get a proton decay rate from \eq{GamM}.  Using the bound $\tau > 1.6\, (0.77)\times 10^{34}$~yr, for $e^+$ $(\mu^+)$, at 90\% confidence level \cite{ParticleDataGroup:2022pth} we find\footnote{PDG 2022 also cites an updated bound, stronger by 3/2, which constrains $\Lambda$ at the same level.}
\be
\Lambda \gsim 2\times 10^{11} \left(\frac{g_N}{10^{-25}}\right)^{1/3}~{\rm GeV}\,.
\label{LamEarth}    
\ee

Next, let us consider the decay of nucleons in the Sun, which would lead to an anomalous flux of $\ord{10~{\rm MeV}}$ neutrinos.  The Super-Kamiokande (SK) experiment has searched for such a flux, from possible baryon number violation processes in the Sun \cite{Super-Kamiokande:2012tld}, mediated by monopoles 
\cite{Rubakov:1981rg,Rubakov:1982fp,Callan:1982ah,Callan:1982au} associated with grand unified theories \cite{tHooft:1974kcl,Polyakov:1974ek}.  

The SK search, with 176 kton-yr of exposure \cite{Super-Kamiokande:2012tld}, focused on a $\pi^+$ final state in proton decay, leading to $\pi^+ \to \mu^+ \nu_\mu \to e^+ {\bar \nu_\mu} \nu_\mu \nu_e$.  Since in our simple example we do not have direct proton decay into $\pi^+$, we consider $p\to e^+ \eta$ where $\eta \to \pi^+\pi^-\pi^0$ and $\eta \to \pi^+\pi^- \gamma$ together have a branching fraction ${\rm Br}(\eta\to \pi^+ \ldots)\approx 27\%$ \cite{ParticleDataGroup:2022pth}.  We can find the $\phi$ profile in the Sun from 
\be
\phi(r_0) = -\frac{g_N}{2 m_N}\int_0^{R_\odot}\!\!\!  dr\, r^2\, \rho(r)
\int_{-1}^{+1} \!\!\! dx\, \frac{e^{-m_\phi |\vec{r}-\vec{r}_0|}}{|\vec{r}-\vec{r}_0|}\,,
\label{phir0}
\ee
where $r_0=|\vec{r}_0|$ is the radial distance from the center of the Sun.  We adopt the BP2004 model of  Ref.~\cite{Bahcall:2004fg} for the Solar mass density $\rho(r)$ of the Sun as a function of the radial coordinate.\footnote{Numerical data for the BP2004 Solar model are available at:  \url{http://www.sns.ias.edu/~jnb/SNdata/Export/BP2004/bp2004stdmodel.dat}.}

The rate ${\cal R}_{\eta e}$ for $p\to \eta \,e^+$ in the Sun is obtained by $\vev{\phi}\to \phi(r)$ in \eq{GamM}, using 
\be
{\cal R}_{\eta e} = \frac{4 \pi}{m_N} \int_0^{R_\odot} \!\!\! dr\, r^2 
\rho(r)\,\Gamma(r)_{(p\to \eta \,e^+)}\,.
\label{Retae}
\ee

The SK 90\% CL limit on the anomalous Solar neutrino flux is $I_{90}=166.6$~cm$^{-2}$ s$^{-1}$ \cite{Super-Kamiokande:2012tld}.  We adapt the SK analysis and obtain 
\be
{\cal R}_{\eta e} = 
\frac{4 \pi\, d_{\rm AU}^2 I_{90}}{3 {\rm Br}(\eta\to \pi^+ \ldots)\,(1 - a_{\pi^+})}\,,
\label{rate90CL}
\ee
where $d_{\rm AU} \approx 1.5 \times 10^8$~km is the Earth orbital radius and 
$a_{\pi^+} = 0.2$ is the $\pi^+$ absorption probability in the center of the Sun, adopted here for the entire Sun, as a simplifying approximation.  We then find 
\be
\Lambda \gsim 2 \times 10^{10}\left(\frac{g_N}{10^{-25}}\right)^{1/3}~{\rm GeV}\,, 
\label{Sol-bound}
\ee
corresponding to the above SK Solar neutrino flux constraint.

\section{Neutron Star Heating via Nucleon Decay}  

We will next examine how the background for $\phi$ sourced by a neutron star (NS) can lead to enhanced decay rates for nucleons, which will lead to anomalous heating of NSs.  We will take a typical NS mass $M_{\rm NS}\approx 1.5 \, M_\odot$ and radius $R_{\rm NS}\approx 10$~km for illustrative purposes.  This yields a nucleon density $n_N \sim 4\times 10^{38}$~cm$^{-3}$.  We will focus on neutron decay $n\to \pi^- e^+$, where the entire neutron mass is deposited in the NS.  Given that $\ord{\rm 10~MeV}$ neutrino scattering cross section on a nucleon is $\sim 10^{-42}$~cm$^2$ \cite{Formaggio:2012cpf}, the mean free path for released neutrinos is $\ord{\rm 10~m}$ which is much smaller than $R_{\rm NS}$, and hence our assumption for the capture of the entire decay energy is valid.        
   
We will assume a constant density for the NS, in our estimate, given by 
\be
\rho_{\rm NS} = \frac{M_{\rm NS}}{(4 \pi/3)R_{\rm NS}^3}\approx 7\times 10^{14}~{\rm g}~{\rm cm}^{-3}\,.
\label{rhoNS}
\ee
For $r<R_{\rm SN}$, we find 
\be
\phi_{\rm NS}(r) \approx -\frac{g_N \,\rho_{\rm NS}}{6\, m_n}\, R_{\rm NS}^2
\left(3 - \frac{r^2}{R_{\rm NS}^2}\right),
\label{phiNS}
\ee
which yields the decay rate in the NS
\be
\Gamma_n^{\rm NS} = 4 \pi\, \frac{\rho_{\rm NS}}{m_n}
\int_0^{R_{\rm NS}} dr \, r^2 \, \Gamma(r)_{(n\to \pi^-\, e^+)}\,.
\label{GamNS} 
\ee

Once the energy deposition from neutron decays and dissipation from the black body radiation balance each other, a steady state is reached for which  
\be
m_n \Gamma_n^{\rm NS} \approx 4 \pi R_{\rm NS}^2 \,\sigma_{\rm SB} T_{\rm NS}^4\,,
\label{SteadyState}
\ee           
where $\sigma_{\rm SB} = \pi^2/60$ is the Stefan-Boltzmann constant and $T_{\rm NS}$ is the NS surface temperature.

The coldest known NS is the pulsar PSR J2144–3933, with $T_{\rm NS} < 42000$~K, from measurements by the {\it Hubble Space Telescope} (HST) \cite{Guillot:2019ugf}.   Estimated to be $\sim 3 \times 10^8$~yr old, this NS is expected to have cooled  to  $T_{\rm NS}\sim \ord{100~{\rm K}}$, in the absence of heating \cite{Yakovlev:2004iq}.  The HST bound on $T_{\rm NS}$ implies 
\be
\Lambda \gsim 7 \times 10^{11}\left(\frac{g_N}{10^{-25}}\right)^{1/3}~{\rm GeV}\,,
\label{HST}
\ee
in the context of our model.  Potential improvements of the above bound may be possible with future data from the James Webb Space Telescope; see for example Refs.~\cite{Chatterjee:2022dhp,Raj:2024kjq}.

Before ending this article, we would like to add that $\phi$ could in principle be ultralight DM, with modest extensions of the model outlined here.  This could involve coupling $\phi$ to electrons which would lead to its misalignment by thermal processes in the early Universe.  At late times, the oscillating $\phi$ would then behave like cold DM.  Such a setup could lead to time varying effects in nucleon decays that were discussed above.  For more details regarding this possibility, see Ref.~\cite{Davoudiasl:2023peu}.   

%
%
%

\section*{Acknowledgments}

We would like to thank the organizers of the EW Moriond 2024, which provided a pleasant venue for interactions and many interesting discussions.  The work of the author is supported by the US Department of Energy, under Grant Contract DE-SC0012704.   

%


\end{document}